\documentclass[]{achemso}
\usepackage[version=3]{mhchem} 
\usepackage[T1]{fontenc}       

\usepackage{amssymb}
\usepackage{amsmath}

\usepackage{color}

\usepackage[normalem]{ulem}
\usepackage{comment}

\author{Tianyu Su}
\affiliation[A]
{Department of Materials Science and Engineering, The Grainger College of Engineering, University of Illinois at Urbana-Champaign, 1304 W. Green Street, Urbana, Illinois 61801, United States}

\author{Brian J. Blankenau}
\affiliation[B]
{Department of Mechanical Science and Engineering, The Grainger College of Engineering, University of Illinois at Urbana-Champaign, 1206 W. Green Street, Urbana, Illinois 61801, United States}

\author{Namhoon Kim}
\affiliation[B]
{Department of Mechanical Science and Engineering, The Grainger College of Engineering, University of Illinois at Urbana-Champaign, 1206 W. Green Street, Urbana, Illinois 61801, United States}

\author{Kshitij Vijayvargia}
\affiliation[B]
{Department of Mechanical Science and Engineering, The Grainger College of Engineering, University of Illinois at Urbana-Champaign, 1206 W. Green Street, Urbana, Illinois 61801, United States}
\alsoaffiliation[D]
{International Institute for Carbon-Neutral Energy Research (WPI-I2CNER), Kyushu University, Fukuoka 819-0395, Japan}

\author{Petros Sofronis}
\affiliation[B]
{Department of Mechanical Science and Engineering, The Grainger College of Engineering, University of Illinois at Urbana-Champaign, 1206 W. Green Street, Urbana, Illinois 61801, United States}
\alsoaffiliation[D]
{International Institute for Carbon-Neutral Energy Research (WPI-I2CNER), Kyushu University, Fukuoka 819-0395, Japan}

\author{Jessica A. Krogstad}
\affiliation[A]
{Department of Materials Science and Engineering, The Grainger College of Engineering, University of Illinois at Urbana-Champaign, 1304 W. Green Street, Urbana, Illinois 61801, United States}
\alsoaffiliation[C]
{Materials Research Laboratory, University of Illinois at Urbana-Champaign, 104 South Goodwin Avenue, Urbana, Illinois 61801, United States}

\author{Elif Ertekin}
\affiliation[B]
{Department of Mechanical Science and Engineering, The Grainger College of Engineering, University of Illinois at Urbana-Champaign, 1206 W. Green Street, Urbana, Illinois 61801, United States}
\alsoaffiliation[C]
{Materials Research Laboratory, University of Illinois at Urbana-Champaign, 104 South Goodwin Avenue, Urbana, Illinois 61801, United States}
\email{ertekin@illinois.edu}

\title{ Short-range order influences H distribution in Fe--Ni--Cr austenitic stainless steels }

\keywords{Hydrogen embrittlement; Short-range order; Density functional theory; Spin cluster expansion; Monte Carlo}

\begin{document}


\newpage
\begin{abstract}

Hydrogen embrittlement (HE) in austenitic stainless steels is advanced by
hydrogen enhanced localized plasticity (HELP), typically accompanied by a transition from homogeneous to localized slip.
Short-range order (SRO) in face-centered cubic (FCC) alloys is known to promote slip planarity, and recent studies suggest that H may amplify this localization behavior linked to inherent SRO.
However, the manner in which the introduction of H affects SRO properties and, conversely, the manner that pre-existing SRO may affect H behavior, are not fully understood.  
In this work, a spin cluster expansion model combined with Monte Carlo simulation is employed to study the interplay between H and SRO in Fe–Ni–Cr alloys. 
Chemical order is quantified using Warren–Cowley SRO parameters, and the model predictions are validated against experimental data. 
We find that the presence of H only slightly alters the intrinsic ordering preference of the Fe--Ni--Cr alloys.
As temperature decreases and the alloy evolves from disordered to ordered thermodynamic states, distinct H–metal correlations emerge. 
In particular, H–Ni and H–Cr pairs exhibit stronger ordering tendencies than H–Fe pairs, suggesting a selective affinity of H for certain atomic environments. 
On the other hand, we also find that compared to random alloys, when pre-existing SRO is present, it significantly affects the resulting H distribution by promoting local H enrichment in SRO domains. 
Such SRO-driven local H accumulation may facilitate slip localization and contribute to the early onset of embrittlement. 
These findings provide thermodynamic and structural insights into the interaction between H and SRO in austenitic stainless steels, highlighting possible implications on how the interaction between HELP and SRO brings about hydrogen embrittlement in austenitic stainless steels.
\end{abstract}

\newpage 
\section{Introduction}

Austenitic stainless steels, despite being among the most compatible systems with hydrogen and therefore widely used in hydrogen energy applications, still suffer from embrittlement~\cite{san2012hydrogen,martin2019enumeration,abraham1995hydrogen,whiteman1965hydrogen}.
Hydrogen facilitates the development of dislocation microstructures in a wide range of metals and alloys~\cite{martin2011interpreting,martin2012hydrogen,martin2013microstructural,wang2017influence,wang2018hydrogen,wang2019assessment,wang2019failure} through hydrogen-induced shielding of dislocation/defect interactions~\cite{birnbaum1994hydrogen,sofronis1995mechanics}, a phenomenon known as hydrogen-enhanced localized plasticity (HELP)~\cite{beachem1972new,abraham1995hydrogen,robertson2015hydrogen,martin2019enumeration}.
HELP is the underlying mechanism of hydrogen embrittlement of austenitic steels, and its manifestation is fracture advanced by plastic flow localization~\cite{nibur2009role,jackson2016effects}. 
It is worth noting that while hydrogen enhances slip planarity, localized slip can also occur in austenitic steels even in the absence of hydrogen~\cite{san2012hydrogen,ulmer1991hydrogen,nibur2009role}.
Another possible explanation of austenitic stainless steel embrittlement is martensite formation, but it is well known that stable austenitic steels are also embrittled by hydrogen~\cite{san2012hydrogen,michler2012hydrogen}.  
 
As has been mentioned, embrittlement due to HELP is still challenging to explain, and this highlights the need for a deeper mechanistic understanding on how HELP brings about fracture through slip localization. 
The hydrogen-induced reduction of the stalking fault energy has been proposed as the underlying reason for slip planarity~\cite{gibbs2020stacking}.  
However, as Ferreira, Robertson, and Birnbaum have argued, it is very unlikely that the small magnitude of the stalking fault energy in these steels is the reason for slip planarity~\cite{ferreira1996influence}.
In addition, there is a range of alloying elements with seemingly contradictory effects on the mechanical response of austenitic steels.
For example, Cr improves H compatibility despite reducing SFE~\cite{symons1997hydrogen}, and the inclusion of N in Nitronic 40, sometimes referred to by its composition as 21Cr-6Ni-9Mn, can lead to ductility losses despite being a strong austenite stabilizer~\cite{odegard1976effect} 
Similarly, Al increases SFE but stabilizes the ferrite phase at annealing temperatures~\cite{jackson1984high,medvedeva2014first}, while Mn increases thermodynamic stability against martensite but decreases SFE~\cite{klueh1988manganese,vitos2006alloying}. 

As numerous studies exploring the compositional dependence of HE behavior have demonstrated~\cite{san2008effects,zhang2008effect,lee2021comparative}, new metrics that account for intrinsic chemical composition are needed to understand how HELP advances fracture in austenitic stainless steels.  
Recently, the role of short-range order (SRO) on slip planarity has gained increasing attention. 
In technologically relevant materials like high entropy alloys, SRO is known to affect SFE, modify the energy landscape for dislocation motion, and promote deformation localization via glide plane softening mechanisms~\cite{ding2018tunable,li2019strengthening,zhang2020short}. 
Experimental studies suggest that H can modify the degree of SRO in austenitic stainless steels~\cite{flanagan1986hydrogen,kim2015effect,kim2019role}, and that SRO is linked to increased slip planarity and a transition from homogeneous to localized deformation~\cite{gerold1989origin}.

Meanwhile, other studies suggest that H enhances the shear localization process caused by pre-existing SRO~\cite{koyama2021potential,gavriljuk2024hydrogen}.
Variations in H solubility across different SRO domains could lead to local H enrichment, which might accelerate dislocation velocity by shielding dislocation interactions according to the HELP mechanism~\cite{birnbaum1994hydrogen,sofronis2001hydrogen}.
These findings highlight the complex relationship between SRO and HELP.

To systematically investigate the interplay between H, HELP, and SRO in austenitic stainless steels, in this work we develop a spin cluster expansion (CE) model that enables quantitative evaluation of SRO in the presence of H.
The CE method provides an approximate yet efficient method to obtain the configurational energy of multicomponent systems~\cite{Sanchez1984,wolverton1994cluster}.
Monte Carlo simulations are performed to sample the alloy configuration across a range of temperatures and compositions. 
The Warren-Cowley SRO parameters~\cite{cowley1965shortrange}, quantifying the chemical order of alloys, are validated against available experimental measurements.
We find that the introduction of H into the alloys only slightly alters the intrinsic thermodynamic driving forces for chemical ordering.
As temperature decreases and intrinsic chemical ordering strengthens, H–metal correlations likewise become more pronounced.
In particular, clustering of H atoms in specific local chemical environments is observed as SRO structures develop in alloys.
This segregation behavior may affect the mechanical properties of alloys due to the hydrogen shielding effect. 
Independent of thermodynamic considerations, the presence of pre-existing SRO in the alloy has a strong influence on the resulting spatial distribution of hydrogen. 
These findings provide atomistic insight into the interactions of the HELP mechanism with SRO nanodoamins, and highlight the importance of chemical ordering for the design of H-resistant alloys.

\section{Methods}

\subsection{First-principles data generation}

Density functional theory (DFT) calculations with spin polarization were conducted using the Projected Augmented Wave (PAW) method~\cite{kresse1999ultrasoft}, implemented in the Vienna \textit{Ab-initio} Simulation Package (VASP)~\cite{kresse1993ab,kresse1996efficient}.
The exchange-correlation interactions were treated using the Perdew-Burke-Ernzerhof (PBE) functional\cite{perdew1996generalized}, and PAW-PBE pseudopotentials were applied with frozen semi-core states. 
The valence electron configurations for Fe, Ni, Cr, and H were specified as [Ar]3d$^7$4s$^1$, [Ar]3d$^9$4s$^1$ [Ar]3d$^5$4s$^1$, and 1s$^1$, respectively. 
The plane wave basis set was truncated at 500 eV.
Fermi-level smearing was applied using the first-order Methfessel-Paxton method, with a smearing width of 0.05 eV.
A $k$-point mesh density of 2400 $k$-points per reciprocal atom was adopted, corresponding to a $11 \times 11 \times 11$ Monkhorst-Pack $\Gamma$-centered mesh for a single atom FCC unit cell. 
Convergence tests were performed to ensure the $k$-point sampling maintained an accuracy within 0.4 meV/atom.
Structural relaxations were carried out until the total energy reached a precision of $10^{-6}$ eV/cell, while atomic forces were minimized below 0.02 eV/\AA. 
Different magnetic states are initialized for the same structures to introduce magnetic degrees of freedom into the system, following the same approach as described in our previous work~\cite{su2024first,su2025nitrogen}. 

The FCC Fe--Ni--Cr alloy structures without H are generated automatically by the Alloy Theoretic Automated Toolkit (ATAT)~\cite{van2002alloy} via a variance reduction scheme. 
To incorporate H into the system, we used 2$\times$2$\times$2 special quasi-random structures (SQS) with 32 metal atoms of varying compositions, generated using the \texttt{mcsqs} code~\cite{van2013efficient}.
Different degrees of H incorporation, i.e., 1, 2, 4, 8, or 16 H atoms, are inserted randomly into the octahedral interstitial sites of the SQS structures, as H prefers to occupy the octahedral site in FCC Fe--Ni--Cr alloys~\cite{zhou2022fe}.
Detailed alloy compositions and a representative SQS structure containing H are provided in the supplementary information (SI) Figure S1(a,b), respectively.
Hydrides (FeH, NiH, and CrH with cubic B1 structure) are also included in the dataset.
In total, the complete DFT dataset consists of 533 pure alloy structures and 273 H-containing structures.

\subsection{Cluster expansion -- Monte Carlo simulation}

Building on our previous work~\cite{kim2022multisublattice,su2024first}, we developed a spin cluster expansion (CE) framework for the Fe--Ni--Cr--H system using our in-house code, the CLAMM toolkit~\cite{blankenau2025clamm}.
The importance of magnetic interactions in transition metal alloys has been emphasized in prior studies, which demonstrated that Cr ordering is strongly influenced by magnetism~\cite{niu2015spin,walsh2021magnetically,su2024first}.
To facilitate such investigations, we developed the open-source software suite CLAMM, which enables simulations of thermodynamic, magnetic, and structural properties of complex alloys and magnetic materials within the spin CE framework. 
The corresponding Hamiltonian is expressed as a function of both chemical and magnetic interactions:
\begin{equation}
    E_{CE}(\vec{\sigma}) = \sum_{\alpha} J_{\alpha} \Theta_{\alpha} (\vec{\sigma}) + \sum_{\beta} \sum_{\langle i,j \rangle} J_{\beta} S_{i} S_{j} \hspace{1em}. \label{ce}
\end{equation}
The first term describes the chemical configurational energy, where $J_{\alpha}$ is the effective cluster interaction (ECI) of a cluster $\alpha$ and $\Theta_{\alpha}(\vec{\sigma})$ denotes its occurrence in a given alloy configuration $\vec{\sigma}$. 
The cluster $\alpha$ is defined on the crystal lattice with a specific structural motif and atomic decoration, and $\vec{\sigma}$ is the sequence of atomic species $\sigma_i$ occupying lattice sites $i$.
The second term represents magnetic exchange interactions between spin pairs $\langle i,j \rangle$, with $J_{\beta}$ denoting the exchange constant for spin dimer $\beta$ and $S_i$, $S_j$ representing the spins.
More details about this formalism can be found in our previous work~\cite{su2024first,su2025nitrogen,blankenau2025clamm}.

For sampling, we implemented the lattice Monte Carlo (MC) method in the canonical ensemble.
Each MC step involved atom swaps and spin-flip trials with equal probabilities, enabling simultaneous configurational and magnetic equilibration.
Temperatures ranging from 500 K to 1500 K were explored with intervals of 100 K.
Initial disordered configurations were equilibrated at the highest temperature, then cooled down and equilibrated at each sequential temperature to evaluate thermodynamic quantities of interest.
The Metropolis-Hastings algorithm~\cite{metropolis_1953} was employed to sample configurations from the Boltzmann distribution.
We used a 10$\times$10$\times$10 conventional FCC supercell, containing 4000 metal atoms with varying amounts of H atoms incorporated, to exclude finite-size effects and ensure energy convergence within 0.1 meV/atom. 
At each target temperature, 2000 MC steps per atom were performed for equilibration, followed by 6000 passes for the evaluation of thermodynamic properties.
Convergence tests confirmed that the number of steps was sufficient for the system to reach equilibrium also to within 0.1 meV/atom. 

As a measure or SRO, we use the Warren-Cowley  parameter defined as
\begin{equation}
    \alpha_l^{AB} = 1 - \frac{P_l^{AB}}{C_AC_B} = 1 - \frac{p_{l,A}^B}{C_B} \hspace{0.5em}, 
\label{WC_SRO}
\end{equation} 
where $P_l^{AB}$ represents the probability of finding $AB$ pairs in the $l$-th neighbor shell, and $p_{l,A}^B$ = $P_l^{AB}$/$C_A$ denotes the conditional probability of finding an atom $B$ in the $l$-th coordination shell around an atom $A$. 
Here $C_A$ and $C_B$ are the concentrations of species $A$ and $B$, respectively. 
In the case of a random solution with no chemical correlation, $\alpha$ vanishes since $P_l^{AB}$ = $C_AC_B$. 
A positive $\alpha$ indicates a tendency toward like-pairs ($AA$ and $BB$ clustering), whereas a negative value of $\alpha$ suggests an ordering preference for unlike pairs ($AB$ ordering).

While the original Warren–Cowley formalism was primarily applied to metallic alloys, it can be extended to H–metal SRO defined as 
\begin{equation}
    \alpha_l^{H-M} = 1 - \frac{n_{l}^M}{K_{l}C_M} \hspace{0.5em}, 
\label{H_metal_SRO}
\end{equation} 
where $K_l$ and $n_{l}^M$ are the total number of metal atoms (coordination number) and the number of metal atoms M in the $l_{th}$ shell of the H atom, respectively. 
The concentration of metal species M is denoted as $C_M$.
Note that $\frac{n_{l}^M}{K_{l}C_M}$ is the conditional probability of finding metal atom $M$ in the $l$-th coordination shell of an H atom.
These H–metal SRO parameters can go from negative values to 1.

\section{Results and Discussion}

\subsection{Construction and validation of the spin CE model}

\begin{figure*}[!hbtp] 
\centering
\includegraphics[width=6in]{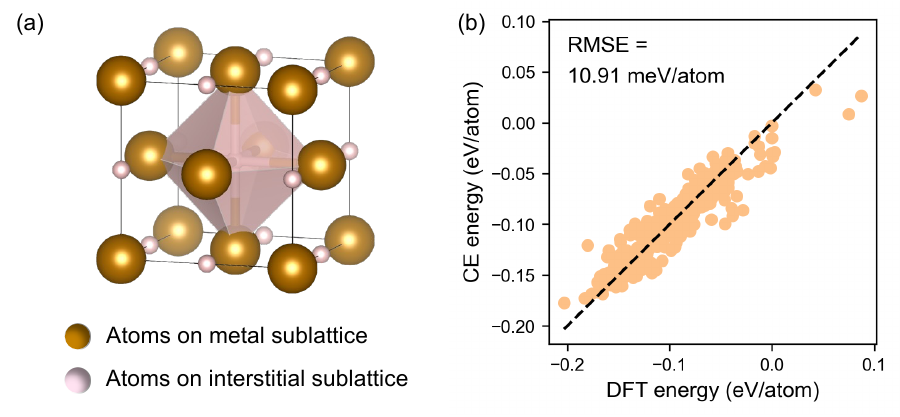}
\caption{
(a) Scheme for interstitial H atoms occupying octahedral sites within the FCC lattice, with the center octahedral interstitial site highlighted in the cell. Two interpenetrating FCC sublattices are present. The yellow atoms label the sites on the metal sublattice, while the pink atoms label the sites on the interstitial sublattice. (b) The comparison between DFT-calculated formation energies and
predictions from the spin CE model for the Fe–Ni–Cr–H dataset. }
\label{fig1} 
\end{figure*}

The Fe--Ni--Cr--H system consists of two interpenetrating FCC sublattices, with one occupied by Fe/Ni/Cr atoms and the other by H atoms, as illustrated in Figure~\ref{fig1}(a). 
The spin CE model, containing both inter- and intra-sublattice interactions, was fitted to the Fe--Ni--Cr--H DFT dataset (see Method section).
The optimal CE model consists of 15 chemical dimers, 34 chemical trimers, 3 chemical quadrumers, and 3 spin dimers. 
The complete list of input clusters and their corresponding ECIs are provided in SI Table S1 and Figure S2.
Using the Least Absolute Shrinkage and Selection Operator (LASSO)~\cite{tibshirani1996regression}, the model selects 105 non-zero ECIs, including decoration,  from the full feature set. 
The comparison between the DFT-calculated energies and the CE-predicted energies is shown in Figure~\ref{fig1}(b), indicating  accuracy in capturing the energetics of the model system.
The root mean square error (RMSE) of the spin CE model is 10.91 meV/atom, similar to previous CE studies of Fe--Ni--Cr based alloys~\cite{wrobel2015phase,su2024first}.

To validate the accuracy of the spin CE model, benchmark tests were conducted against available experimental data of austenitic stainless steels. 
These benchmarks, containing experimental measurements of Warren-Cowley SRO parameters, order-disorder transition temperatures, and Curie temperatures, have been reported in our previous work~\cite{su2024first}. 
A detailed comparison between the spin CE model predictions and experimental results, including SRO parameters, order-disorder transition temperatures, and Curie temperatures of various metals and alloys, is presented in SI Figure S3.
Overall, the spin CE model demonstrates good agreement with the experimental benchmarks, confirming its reliability for describing SRO and magnetic behavior in the alloy system.
However, to our knowledge, experimental data for SRO parameters of H-containing alloys are not available.  
This makes it difficult to directly compare the model to experiments in reality. 

\subsection{The effect of H and Cr on SRO}

\begin{figure}[!hbtp] 
\centering
\includegraphics[width=6in]{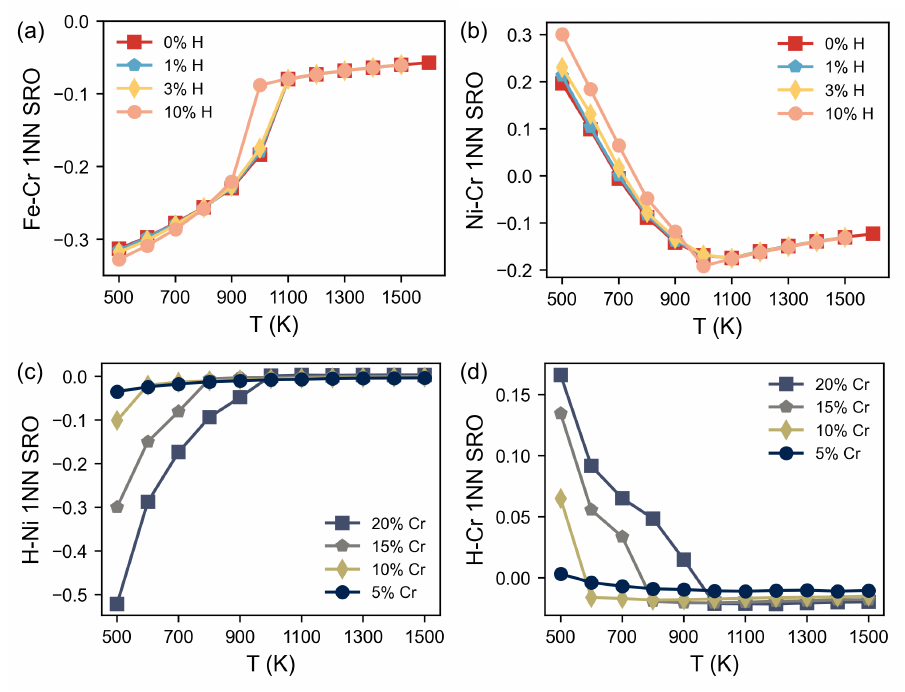}
\caption{
(a,b) The effect of H on Fe–Cr and Ni–Cr 1NN SRO of the Fe$_{70}$Ni$_{10}$Cr$_{20}$ alloy. The H concentration incorporated in the alloy varies from 0 at.\% to 10 at.\%. (c,d) The effect of Cr on H–Ni and H–Cr 1NN SRO for different alloys with 10 at.\% H atoms. The Ni concentration of the alloy is fixed at 10 at.\% while the Cr concentration varies from 5 at.\% to 20 at.\%. 
All panels illustrate thermodynamic ordering tendencies.}
\label{fig2}
\end{figure}

Based on the spin CE model constructed above, we performed MC simulations for a representative austenitic stainless steel Fe$_{70}$Ni$_{10}$Cr$_{20}$.
Various concentrations of H atoms, ranging from 0 to 10 at.\%, were introduced into the simulation cell.
Figure~\ref{fig2}(a,b) show the effects of H on the 1NN SRO parameters for Fe–Cr and Ni–Cr pairs, respectively.
The Fe–Ni 1NN SRO parameter is not shown due to its predicted negligible magnitude~\cite{su2024first}.
As H concentration increases from 0 to 3 at.\%, no significant changes in Fe–Cr or Ni–Cr SRO parameters can be observed (the curves nearly lie on top of each other). 
Only when H concentration reaches 10 at.\%, the order-disorder temperature for the Fe–Cr 1NN pair decreases slightly by 100 K, while the Ni–Cr SRO becomes somewhat more positive at low temperatures.
These results suggest that H only weakly disrupts the intrinsic thermodynamic ordering tendency between Fe–Cr and Ni–Cr pairs, possibly due to the interaction between H and metal atoms as shown in SI Figure S4. 
And because most of the H–metal interactions are small compared to metal-metal interactions, the influence of H on the intrinsic SRO is limited.
This finding contrasts with a previous experimental observation where H seems to induce the formation of SRO during deformation in austenitic stainless steels~\cite{kim2019role}.
The diffuse scattering observed after deformation was attributed to the formation of SRO, but alternatively may have arisen from the increasing population of planar defects~\cite{kung2025differentiating}.
From a thermodynamic perspective, the results from the spin CE model do not support a strong effect of H on promoting SRO.

In contrast, Cr species have a strong influence on SRO due to its complex magnetic properties, as demonstrated in previous studies~\cite{niu2015spin,walsh2021magnetically,su2024first}.
The effect of Cr concentration on H-related SRO is shown in Figure~\ref{fig2}(c,d).
Note that the H–Fe 1NN SRO parameter is not shown because it is close to zero.
The alloy composition is chosen such that the Ni concentration is fixed at 10 at.\% while the Cr content varies from 5 at.\% to 20 at.\%.
The rest of the metal sublattice is occupied by Fe, and the H concentration is maintained at 10 at.\% (H/M atomic ratio) throughout, which corresponds to 0.18 wt.\%.
These are compositions that are typical of austenitic stainless steel.
When Cr concentration is small (5 at.\%), both H–Ni and H–Cr 1NN SRO parameters remain close to zero across the entire temperature range, indicating little preference for H to associate with these substitutional metal atoms.
As Cr concentration increases from 5 at.\% to 20 at.\%, the H–Ni 1NN SRO becomes increasingly negative, while the H–Cr 1NN SRO becomes more positive.
The growing magnitude of these two SRO parameters reflects the strong preference for H–Ni pairs and avoidance of H–Cr pairs.
In austenitic stainless steels, this implies that higher Cr concentrations enhance the likelihood of H occupying interstitial sites near Ni atoms.

This behavior is related to the evolution of intrinsic SRO structures in the alloy as the Cr concentration changes (described further below and in SI Figure S4).
The distribution of H atoms in a representative high Cr composition, Fe$_{70}$Ni$_{10}$Cr$_{20}$, is further explored in the next section. 

\subsection{Changes in H distribution accompanying SRO development}

\begin{figure}[!hbtp] 
\centering
\includegraphics[width=6in]{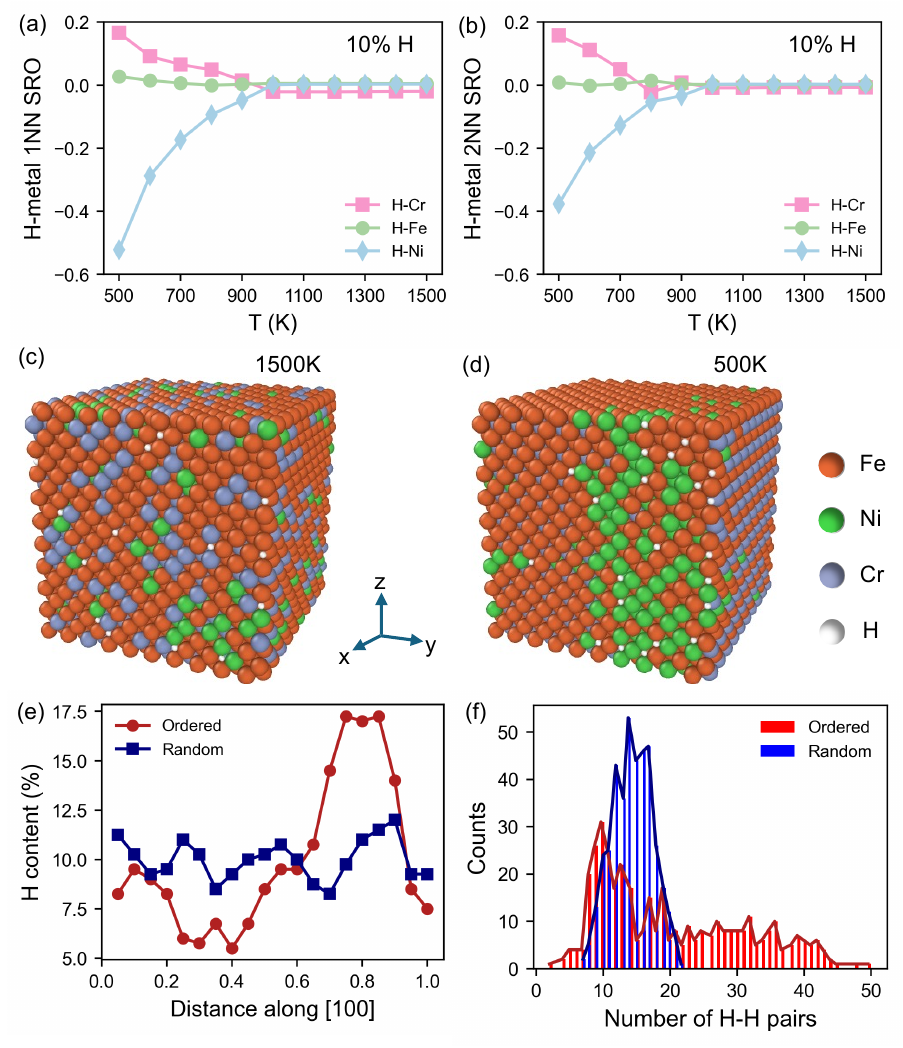}
\caption{Thermodynamic ordering tendencies characterized by (a) H–metal 1NN, and (b) H–metal 2NN SRO parameters for the Fe$_{70}$Ni$_{10}$Cr$_{20}$ alloy incorporated with 10\% H atoms. 
(c,d) The corresponding representative MC snapshots generated at 1500~K and 500~K, visualized using the OVITO software~\cite{stukowski2009visualization}. Fe, Ni, Cr, and H atoms are marked as red, green, blue, and white, respectively. (e) H concentration for ordered (500 K) and random (1500 K) configurations plotted here along [100] direction to highlight the compositional variation. (f) The distribution of H atoms as a function of the number of H–H pairs within two FCC unit cells.
}
\label{fig3}
\end{figure} 

For the Fe$_{70}$Ni$_{10}$Cr$_{20}$ alloy with 10 at.\% H, the 1NN and 2NN H–metal SRO parameters are shown in Figure~\ref{fig3}(a,b).
At high temperatures, all H–metal pairs exhibit negligible SRO parameters, indicating no significant ordering preference.
As the temperature decreases to around 1000 K, the magnitudes of H–Ni and H–Cr SRO parameters increase, marking the emergence of H-related ordering.
Notably, this onset coincides with the development of chemical order among the metal atoms in the alloy (see Figure~\ref{fig2}(a)). 
Compared to the negligible H–Fe SRO, the negative H–Ni 1NN and 2NN SRO parameters indicate the strong preference for H to occupy interstitial sites near Ni atoms.
On the contrary, the positive H–Cr SRO parameters reflect an avoidance between H and Cr atoms.

To investigate the spatial distribution of H in the alloy, representative MC snapshots are shown in Figure~\ref{fig3}(c,d).
The high temperature configuration generated at 1500 K (Figure~\ref{fig3}(c)) appears nearly random, while the low temperature configuration at 500 K (Figure~\ref{fig3}(d)) exhibits stronger chemical ordering, including Fe–Ni and Fe–Cr rich domains.  
As the ordering becomes more pronounced, as shown in our previous work \cite{su2024first},  these domains show structural features that resemble Fe$_3$Cr and FeNi intermetallic structures.
Interestingly, the H distribution seems to be influenced by this intrinsic metal atom ordering.
Figure~\ref{fig3}(e) compares the H concentration profiles of these two configurations 
At 500 K (ordered alloy), the H concentration profile reveals localized H segregation and depletion, with peak contents around 17.5 at.\% and minimum contents around 5 at.\%.
The H-rich region corresponds to the Fe–Ni rich domain, and the H-depleted region to the Fe–Cr rich domain. 
In comparison, the H profile at 1500 K (random alloy) is much flatter with only small fluctuations, reflecting a relatively uniform distribution.

To quantify this effect, Figure~\ref{fig3}(f) shows the distribution of H atoms as a function of the number of H neighbors within two FCC unit cells of each H atom.
A higher number of H neighbors indicates stronger local clustering behavior.
In the random alloy, the H–H pair count resembles a normal distribution.
However, in the ordered configuration, a long tail appears, indicating that more H atoms have an increased number of H neighbors. 
This suggests a pronounced H clustering tendency in the presence of chemical order.
The concentration profiles of metal atoms along with H are shown in SI Figure S5. 
In the random alloy at 1500 K, only small fluctuations appear in these profiles, with no obvious correlation between H and metal atoms.
In contrast, for the ordered configuration, the observed H clustering regularly coincides with local increases in Ni concentration and decreases in Fe or Cr concentration. 
This behavior indicates a potential link between H segregation and chemical order in Fe--Ni--Cr alloys.


In the spin CE model, H–H interactions are found to be weak, as indicated by the vanishing H-H ECIs during LASSO fitting. 
Instead, H–metal interactions play a dominant role in driving H segregation into SRO domains in Fe--Ni--Cr alloys.
Although H–metal interactions are generally weaker than metal–metal interactions, a few of the key chemical interactions between H and metal atoms (shown in SI Figure S4) are comparable in magnitude to metal–metal interactions.  
As a result, certain SRO structures can potentially attract (or repel) additional H atoms due to these favorable (or repulsive) H–metal interactions. 
As shown in our previous work, the intrinsic ordering tendency of Fe--Ni--Cr alloys suppresses the formation of Cr–Cr and Ni–Cr 1NN pairs while favoring Fe–Cr 1NN pairs~\cite{su2024first}. 
This tendency makes it difficult for H to occupy Fe–Cr-rich SRO domains because of the repulsive interactions (e.g., cluster J containing Fe–Cr 1NN pairs in SI Figure S4).  
Instead, the synergistic effect of metal-metal interactions and attractive H–Ni interactions (e.g., cluster E and F in SI Figure S4) facilitates a stronger tendency for H to segregate in Ni-rich SRO domains. 

\subsection{Distribution of H in fixed-lattice alloys}

\begin{figure*}[!hbtp] 
\centering
\includegraphics[width=6in]{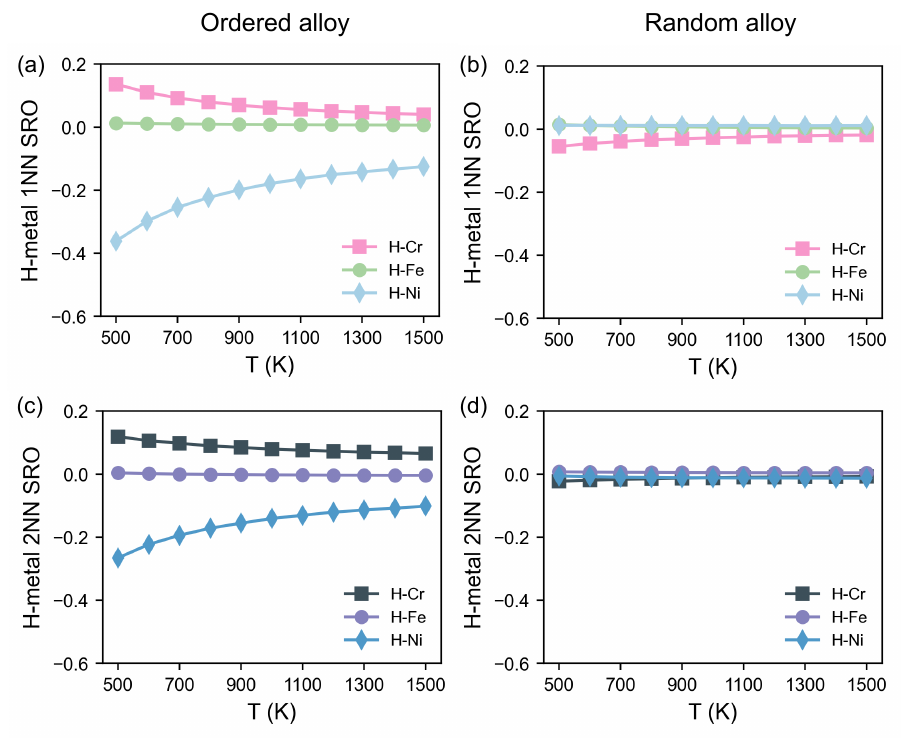}
\caption{
(a,b) H–metal 1 NN SRO parameter as a function of temperature for the fixed-lattice Fe$_{70}$Ni$_{10}$Cr$_{20}$ alloy in ordered and random states, respectively. (c,d) H–metal 2 NN SRO parameter as a function of temperature for the fixed-lattice Fe$_{70}$Ni$_{10}$Cr$_{20}$ alloy in ordered and random states, respectively. The H concentration is 10 at.\% in the alloys. 
}
\label{fig4} 
\end{figure*} 

Since H tends to cluster with Ni in the thermal equilibrium state, the alloy configuration itself may show a significant influence on H distribution. 
Thus far, the results presented have considered a hypothetical fully equilibrated scenario in which all atoms, including metals, are free to move and rearrange in response to H incorporation. 
In practice, however, metal atoms diffuse much more slowly than hydrogen. 
It is therefore more realistic to examine H distributions within fixed metal configurations, focusing on limiting cases of ordered versus random alloy structures.
Figure~\ref{fig4} compares H–metal SRO parameters in ordered and random Fe$_{70}$Ni$_{10}$Cr$_{20}$ alloys with 10\% H incorporation. 
The ordered configuration is sampled at 500 K, while the random configuration is taken from 1500 K MC simulations of the alloy without H. These results illustrate how pre-existing SRO, regardless of its origin, can influence H segregation tendencies.

For the ordered case, H–Ni 1NN and 2NN SRO develops as temperature decreases. 
This is consistent with the trend shown in Figure~\ref{fig3}(a,b), meaning H accumulation in Ni-rich SRO domains. 
The H–Fe SRO parameters are always close to zero, meaning no preference between H and Fe.
The H–Cr SRO parameters become more positive as temperature decreases, meaning that H–Cr pairs are unfavored.
The random alloy, on the other hand, gives small SRO parameters for all three H–metals pairs, leading to random H distribution. 

These results indicate that alloys with pre-existing SRO domains will tend to have a different H distribution compared to random alloys.
In random alloys, it is less likely that H atoms segregate in the matrix.
However, in alloys where SRO domains are observed, H can be accommodated within the Ni-rich SRO domains, thus leading to local H accumulation. 
This finding mainly reflects chemical interactions between metal and H.  
Strain fields from SRO domains also influence H accumulation, and our model only implicitly includes lattice strain via local relaxation of structures in the training dataset of the spin CE. 
It is important to note that additional strain effects on H distribution are therefore not captured here.


\subsection{The relationship between SRO and HE}

The spin CE approach adopted here describes the energetics between H and metal atoms in terms of ECIs, to characterize the interaction between H and SRO. 
As discussed above, the clusters consisting of H–Ni pairs tend to have negative ECIs, while the clusters with H–Fe and Fe–Cr pairs have positive ECIs.
However, the magnitude of the interactions between H and metal atoms is usually smaller than those of the metal-metal interactions themselves.
This explains the MC results that H does not change or disrupt the intrinsic SRO between metal atoms themselves, but instead ``follows suit'' and arranges itself in favorable ways given the degree of metal SRO already present.  
Consequently, chemical order between metal atoms can affect the resulting distribution of H.
In particular, certain SRO domains show a strong affinity for H atoms, leading to local H accumulation.

Pre-existing SRO in the alloys, possibly introduced by thermomechanical processing, might affect the distribution, diffusion, and accumulation of H, thereby influencing the HE processes.
As demonstrated above, Fe–Ni–rich SRO domains strongly attract H, primarily through favorable H–Ni interactions, leading to elevated local H concentrations.
In addition to this chemical effect, the mechanical effect of lattice mismatch between the ordered domain and the surrounding matrix also attracts H. 
In reality, Fe–Ni alloys have smaller lattice parameters compared to Fe--Ni--Cr alloys~\cite{beskrovni1999effect,reed1969lattice}.
Therefore, the Fe–Ni-rich SRO domains, with lattice constant smaller than that of the surrounding matrix, can further attract H atoms due to the tensile stress field. 
Consequently, the accumulation of H atoms inside the SRO domain will reduce the lattice mismatch. 
Recent work~\cite{vijay2025on} demonstrates that the reduction of SRO misfit strain in the presence of H leads to glide plane softening, which in turn triggers the localization of plastic flow in bands of intense shear. 
Thus, the interaction between SRO and H can provide a mechanistic pathway toward understanding the effects of composition on the manifestation of HELP through plastic flow localization.

Given the relatively slow diffusion kinetics of metal atoms compared to H, it is unlikely that H directly induces SRO under ambient conditions. 
Recent studies by~\citeauthor{koyama2021potential} and~\citeauthor{gavriljuk2024hydrogen} stated that pre-existing SRO may play an active role in promoting HE~\cite{koyama2021potential,gavriljuk2024hydrogen}, because, as has been argued, SRO leads to slip localization through the glide plane softening mechanism~\cite{marucco1988effects,gerold1989origin}. 
Moreover, H solubility varies in the submicron volumes 
of distinct SRO regions, and local H enrichment may enhance slip localization through the H shielding effect~\cite{birnbaum1994hydrogen,sofronis2001hydrogen,gavriljuk2024electron} and the weakening of local SRO-induced lattice distortions.
The simulations in this work suggest that the presence of SRO may facilitate slip localization by promoting H segregation which relaxes the local SRO-induced lattice strains, in addition to the previously proposed slip planarity advanced by HELP. 
Hydrogen segregation reduces the lattice mismatch and relaxes the local stress field of SRO, making it easier for dislocations to cut through. 
Consequently, the concerted action of HELP and SRO induces the deformation to localize into bands of intense shear which may result in highly localized fractures~\cite{vijay2025on}, as have been observed experimentally.
On the other hand, tailoring alloy compositions and microstructures to suppress H-attractive SRO formation offers a potential strategy to mitigate these effects.

\section{Conclusions}

In summary, we present a spin cluster expansion model combined with Monte Carlo simulations to investigate the interplay between H and SRO in the Fe--Ni--Cr austenitic stainless steels.
In contrast to some previous experimental studies that have reported H-induced SRO formation in the Fe–Ni–Cr alloy system, the current computational results indicate that H has a limited effect on altering the intrinsic SRO behavior in thermal equilibrium. 
Instead however, the presence of pre-existing chemical order in the alloy, 
particularly SRO domains resembling Fe$_3$Cr and FeNi intermetallics, 
significantly impacts the H distribution. 
In chemically ordered configurations, particularly those with pronounced Fe–Ni SRO domains, H exhibits a strong tendency to segregate near Ni atoms, leading to the formation of locally H-enriched regions. 
In contrast, random alloys display nearly uniform H distribution, underscoring the critical role of atomic ordering in promoting H localization. 
These findings suggest that SRO domains may serve as favorable sites for H accumulation, potentially facilitating the HELP mechanism. 
The interaction between H and SRO offers a plausible atomistic mechanism linking alloy microstructure to H-induced mechanical degradation. 

\section{Acknowledgments}

The authors acknowledge support from the U.S. Department of Energy’s Office of Energy Efficiency and Renewable Energy (EERE) under the Hydrogen and Fuel Cell Technologies Office Award Number DE-EE0008832.
This work was partially supported by the US Department of Energy/National Nuclear Security Administration through the Chicago/DOE Alliance Center, cooperative agreements DE-NA0003975 and DE-NA0004153. 
This work used PSC Bridges-2 at the Pittsburgh Supercomputing Center through allocation MAT220011 from the Advanced Cyberinfrastructure Coordination Ecosystem: Services \& Support (ACCESS) program, which is supported by National Science Foundation grants \#2138259, \#2138286, \#2138307, \#2137603, and \#2138296.

\section{Data Availability}
All data~\cite{H_dataset_zenodo} supporting this work are available at https://doi.org/10.5281/zenodo.16387319.
The cluster expansion and Monte Carlo code are available at https://github.com/ertekin-research-group/CLAMM.

\newpage
\bibliography{mybib.bib}

@article{beachem1972new,
  title={A new model for hydrogen-assisted cracking (hydrogen “embrittlement”)},
  author={Beachem, Cedric D},
  journal={Metallurgical transactions},
  volume={3},
  number={2},
  pages={441--455},
  year={1972},
  publisher={Springer}
}

@article{michler2012hydrogen,
  title={Hydrogen environment embrittlement of stable austenitic steels},
  author={Michler, Thorsten and San Marchi, Chris and Naumann, J{\"o}rg and Weber, Sebastian and Martin, Mauro},
  journal={International journal of hydrogen energy},
  volume={37},
  number={21},
  pages={16231--16246},
  year={2012},
  publisher={Elsevier}
}

@article{nibur2009role,
  title={The role of localized deformation in hydrogen-assisted crack propagation in 21Cr--6Ni--9Mn stainless steel},
  author={Nibur, KA and Somerday, BP and Balch, DK and San Marchi, C},
  journal={Acta Materialia},
  volume={57},
  number={13},
  pages={3795--3809},
  year={2009},
  publisher={Elsevier}
}

@article{jackson2016effects,
  title={Effects of low temperature on hydrogen-assisted crack growth in forged 304L austenitic stainless steel},
  author={Jackson, Heather and San Marchi, Chris and Balch, Dorian and Somerday, Brian and Michael, Joseph},
  journal={Metallurgical and Materials Transactions A},
  volume={47},
  number={8},
  pages={4334--4350},
  year={2016},
  publisher={Springer}
}

@article{ulmer1991hydrogen,
  title={Hydrogen-induced strain localization and failure of austenitic stainless steels at high hydrogen concentrations},
  author={Ulmer, DG and Altstetter, CJ},
  journal={Acta Metallurgica et Materialia},
  volume={39},
  number={6},
  pages={1237--1248},
  year={1991},
  publisher={Elsevier}
}

@article{martin2011interpreting,
  title={Interpreting hydrogen-induced fracture surfaces in terms of deformation processes: A new approach},
  author={Martin, ML and Robertson, IM and Sofronis, P},
  journal={Acta Materialia},
  volume={59},
  number={9},
  pages={3680--3687},
  year={2011},
  publisher={Elsevier}
}

@article{martin2012hydrogen,
  title={Hydrogen-induced intergranular failure in nickel revisited},
  author={Martin, ML and Somerday, BP and Ritchie, RO and Sofronis, P and Robertson, IM},
  journal={Acta Materialia},
  volume={60},
  number={6-7},
  pages={2739--2745},
  year={2012},
  publisher={Elsevier}
}

@article{wang2017influence,
  title={Influence of hydrogen on dislocation self-organization in Ni},
  author={Wang, Shuai and Nagao, Akihide and Edalati, Kaveh and Horita, Zenji and Robertson, Ian M},
  journal={Acta Materialia},
  volume={135},
  pages={96--102},
  year={2017},
  publisher={Elsevier}
}

@article{martin2013microstructural,
  title={A microstructural based understanding of hydrogen-enhanced fatigue of stainless steels},
  author={Martin, ML and Sofronis, P and Robertson, IM and Awane, T and Murakami, Y},
  journal={International journal of fatigue},
  volume={57},
  pages={28--36},
  year={2013},
  publisher={Elsevier}
}

@article{wang2019failure,
  title={On the failure of surface damage to assess the hydrogen-enhanced deformation ahead of crack tip in a cyclically loaded austenitic stainless steel},
  author={Wang, Shuai and Nygren, Kelly E and Nagao, Akihide and Sofronis, Petros and Robertson, Ian M},
  journal={Scripta Materialia},
  volume={166},
  pages={102--106},
  year={2019},
  publisher={Elsevier}
}

@article{wang2018hydrogen,
  title={Hydrogen-modified dislocation structures in a cyclically deformed ferritic-pearlitic low carbon steel},
  author={Wang, Shuai and Nagao, Akihide and Sofronis, Petros and Robertson, Ian M},
  journal={Acta Materialia},
  volume={144},
  pages={164--176},
  year={2018},
  publisher={Elsevier}
}

@article{wang2019assessment,
  title={Assessment of the impact of hydrogen on the stress developed ahead of a fatigue crack},
  author={Wang, Shuai and Nagao, Akihide and Sofronis, Petros and Robertson, Ian M},
  journal={Acta Materialia},
  volume={174},
  pages={181--188},
  year={2019},
  publisher={Elsevier}
}

@article{kung2025differentiating,
  title={Differentiating electron diffuse scattering via 4D-STEM spatial fluctuation and correlation analysis in complex FCC alloys},
  author={Kung, Po-Cheng and Feng, Rui and Liaw, Peter and Zuo, Jian-Min and Krogstad, Jessica},
  journal={Ultramicroscopy},
  pages={114228},
  year={2025},
  publisher={Elsevier}
}

@article{reed1969lattice,
  title={Lattice parameters of martensite and austenite in Fe--Ni alloys},
  author={Reed, RP and Schramm, RE},
  journal={Journal of Applied Physics},
  volume={40},
  number={9},
  pages={3453--3458},
  year={1969},
  publisher={American Institute of Physics}
}

@article{beskrovni1999effect,
  title={Effect of Cr content on the crystal structure and lattice dynamics of FCC Fe--Cr--Ni--N austenitic alloys},
  author={Beskrovni, A and Danilkin, S and Fuess, Hartmut and Jadrowski, E and Neova-Baeva, M and Wieder, T},
  journal={Journal of alloys and compounds},
  volume={291},
  number={1-2},
  pages={262--268},
  year={1999},
  publisher={Elsevier}
}

@article{gibbs2020stacking,
  title={Stacking fault energy based alloy screening for hydrogen compatibility},
  author={Gibbs, Paul J and Hough, Patricia D and Th{\"u}rmer, K and Somerday, Brian P and San Marchi, Chris and Zimmerman, Jonathan A},
  journal={Jom},
  volume={72},
  number={5},
  pages={1982--1992},
  year={2020},
  publisher={Springer}
}

@misc{H_dataset_zenodo,
  author       = {Su, Tianyu},
  title        = {[dataset] H-containing FCC Fe-Ni-Cr alloys dataset},
  year         = {2025},
  publisher    = {Zenodo},
  version      = {v1.0},
  doi          = {10.5281/zenodo.16387319},
  url          = {https://doi.org/10.5281/zenodo.16387319}
}

@article{robertson2015hydrogen,
  title={Hydrogen embrittlement understood},
  author={Robertson, Ian M and Sofronis, Petros and Nagao, Akihide and Martin, May L and Wang, S and Gross, DW and Nygren, KE},
  journal={Metallurgical and materials transactions A},
  volume={46},
  number={6},
  pages={2323--2341},
  year={2015},
  publisher={Springer}
}

@article{vijay2025on,
  title={On hydrogen-induced shear localization in austenitic steels triggered by dislocation interactions with short-range order},
  author={Vijayvargia, K and Hosseini, ZS and Dadfarnia, M and Somerday, BP and Krogstad, JA and Kubota, M and Tsuchiyama, T and Sofronis, P and Aravas, N},
  journal={International Journal of Solids and Structures},
  volume={324},
  pages={113662},
  year={2026},
  publisher={Elsevier}
}

@article{san2012hydrogen,
  title={Hydrogen embrittlement of stainless steels and their welds},
  author={San Marchi, C},
  journal={Gaseous hydrogen embrittlement of materials in energy technologies},
  pages={592--623},
  year={2012},
  publisher={Elsevier}
}

@article{su2025nitrogen,
  title={Nitrogen-related short-range order in Fe-Ni-Cr austenitic stainless steels: first principles and cluster expansion study},
  author={Su, Tianyu and Blankenau, Brian J and Kim, Namhoon and Krogstad, Jessica A and Ertekin, Elif},
  journal={Computational Materials Science},
  volume={260},
  pages={114218},
  year={2025},
  publisher={Elsevier}
}

@article{blankenau2025clamm,
  title={CLAMM: a spin CLuster expansion--Monte Carlo toolkit for Alloys and Magnetic Materials},
  author={Blankenau, Brian and Su, Tianyu and Kim, Namhoon and Ertekin, Elif},
  journal={arXiv preprint arXiv:2506.17800},
  year={2025}
}

@article{whiteman1965hydrogen,
  title={Hydrogen embrittlement of austenitic stainless steel},
  author={Whiteman, MB and Troiano, AR},
  journal={Corrosion},
  volume={21},
  number={2},
  pages={53--56},
  year={1965},
  publisher={Association for Materials Protection and Performance}
}

@article{martin2019enumeration,
  title={Enumeration of the hydrogen-enhanced localized plasticity mechanism for hydrogen embrittlement in structural materials},
  author={Martin, May L and Dadfarnia, Mohsen and Nagao, Akihide and Wang, Shuai and Sofronis, Petros},
  journal={Acta Materialia},
  volume={165},
  pages={734--750},
  year={2019},
  publisher={Elsevier}
}

@article{zhou2022fe,
  title={An Fe--Ni--Cr--H interatomic potential and predictions of hydrogen-affected stacking fault energies in austenitic stainless steels},
  author={Zhou, XW and Nowak, C and Skelton, RS and Foster, ME and Ronevich, JA and San Marchi, C and Sills, RB},
  journal={International Journal of Hydrogen Energy},
  volume={47},
  number={1},
  pages={651--665},
  year={2022},
  publisher={Elsevier}
}

@article{gavriljuk2024electron,
  title={Electron Concept of Hydrogen Embrittlement and Hydrogen-Increased Plasticity of Metals.},
  author={Gavriljuk, VG and Shyvaniuk, VM and Teus, SM},
  journal={Progress in Physics of Metals},
  volume={25},
  number={3},
  year={2024}
}

@article{gavriljuk2024hydrogen,
  title={Hydrogen-dislocation interaction in relation to hydrogen embrittlement and enhanced plasticity of metals},
  author={Gavriljuk, VG and Shyvaniuk, VM and Teus, SM},
  journal={International Journal of Hydrogen Energy},
  volume={50},
  pages={352--360},
  year={2024},
  publisher={Elsevier}
}

@article{su2024first,
  title={First-principles and cluster expansion study of the effect of magnetism on short-range order in Fe--Ni--Cr austenitic stainless steels},
  author={Tianyu Su and Blankenau, Brian J and Kim, Namhoon and Krogstad, Jessica A and Ertekin, Elif},
  journal={Acta Materialia},
  volume={276},
  pages={120088},
  year={2024},
  publisher={Elsevier}
}

@article{stukowski2009visualization,
  title={Visualization and analysis of atomistic simulation data with OVITO--the Open Visualization Tool},
  author={Stukowski, Alexander},
  journal={Modelling and simulation in materials science and engineering},
  volume={18},
  number={1},
  pages={015012},
  year={2009},
  publisher={IOP Publishing}
}

@article{tibshirani1996regression,
  title={Regression shrinkage and selection via the lasso},
  author={Tibshirani, Robert},
  journal={Journal of the Royal Statistical Society Series B: Statistical Methodology},
  volume={58},
  number={1},
  pages={267--288},
  year={1996},
  publisher={Oxford University Press}
}

@article{kim2022multisublattice,
  title={Multisublattice cluster expansion study of short-range ordering in iron-substituted strontium titanate},
  author={Kim, Namhoon and Blankenau, Brian J and Su, Tianyu and Perry, Nicola H and Ertekin, Elif},
  journal={Computational Materials Science},
  volume={202},
  pages={110969},
  year={2022},
  publisher={Elsevier}
}

@article{metropolis_1953,
  author = {Metropolis, Nicholas and Rosenbluth, Arianna W. and Rosenbluth, Marshall N. and Teller, Augusta H. and Teller, Edward},
  month = {06},
  pages = {1087-1092},
  title = {Equation of State Calculations by Fast Computing Machines},
  volume = {21},
  year = {1953},
  journal = {The Journal of Chemical Physics}
}

@article{walsh2021magnetically,
  title={Magnetically driven short-range order can explain anomalous measurements in CrCoNi},
  author={Walsh, Flynn and Asta, Mark and Ritchie, Robert O},
  journal={Proceedings of the National Academy of Sciences},
  volume={118},
  number={13},
  pages={e2020540118},
  year={2021},
  publisher={National Acad Sciences}
}

@article{lee2021comparative,
  title={Comparative study of hydrogen embrittlement resistance between additively and conventionally manufactured 304L austenitic stainless steels},
  author={Lee, Dong-Hyun and Sun, Binhan and Lee, Subin and Ponge, Dirk and J{\"a}gle, Eric A and Raabe, Dierk},
  journal={Materials Science and Engineering: A},
  volume={803},
  pages={140499},
  year={2021},
  publisher={Elsevier}
}

@article{koyama2021potential,
  title={Potential Effects of Short-Range Order on Hydrogen Embrittlement of Stable Austenitic Steels—A Review},
  author={Koyama, Motomichi and Bal, Burak and Canadinc, Dermican and Habib, Kishan and Tsuchiyama, Toshihiro and Tsuzaki, Kaneaki and Akiyama, Eiji},
  journal={Advances in Hydrogen Embrittlement Study},
  pages={1--18},
  year={2021},
  publisher={Springer}
}

@article{abraham1995hydrogen,
  title={Hydrogen-enhanced localization of plasticity in an austenitic stainless steel},
  author={Abraham, Daniel P and Altstetter, Carl J},
  journal={Metallurgical and Materials transactions A},
  volume={26},
  number={11},
  pages={2859--2871},
  year={1995},
  publisher={Springer}
}

@article{marucco1988effects,
  title={Effects of ordering on the properties of Ni-Cr alloys},
  author={Marucco, Alessandra and Nath, Birendra},
  journal={Journal of materials science},
  volume={23},
  number={6},
  pages={2107--2114},
  year={1988},
  publisher={Springer}
}

@article{van2013efficient,
  title={Efficient stochastic generation of special quasirandom structures},
  author={Van de Walle, A and Tiwary, P and De Jong, M and Olmsted, DL and Asta, M and Dick, A and Shin, D and Wang, Yi and Chen, L-Q and Liu, Z-K},
  journal={Calphad},
  volume={42},
  pages={13--18},
  year={2013},
  publisher={Elsevier}
}

@article{van2002alloy,
  title={The alloy theoretic automated toolkit: A user guide},
  author={Van De Walle, Axel and Asta, Mark and Ceder, Gerbrand},
  journal={Calphad},
  volume={26},
  number={4},
  pages={539--553},
  year={2002},
  publisher={Elsevier}
}

@article{perdew1996generalized,
  title={Generalized gradient approximation made simple},
  author={Perdew, John P and Burke, Kieron and Ernzerhof, Matthias},
  journal={Physical review letters},
  volume={77},
  number={18},
  pages={3865},
  year={1996},
  publisher={APS}
}

@article{kresse1999ultrasoft,
  title={From ultrasoft pseudopotentials to the projector augmented-wave method},
  author={Kresse, Georg and Joubert, Daniel},
  journal={Physical review b},
  volume={59},
  number={3},
  pages={1758},
  year={1999},
  publisher={APS}
}

@article{kresse1993ab,
  title={Ab initio molecular dynamics for liquid metals},
  author={Kresse, Georg and Hafner, J{\"u}rgen},
  journal={Physical review B},
  volume={47},
  number={1},
  pages={558},
  year={1993},
  publisher={APS}
}

@article{kresse1996efficient,
  title={Efficient iterative schemes for ab initio total-energy calculations using a plane-wave basis set},
  author={Kresse, Georg and Furthm{\"u}ller, J{\"u}rgen},
  journal={Physical review B},
  volume={54},
  number={16},
  pages={11169},
  year={1996},
  publisher={APS}
}

@article{niu2015spin,
  title={Spin-driven ordering of Cr in the equiatomic high entropy alloy NiFeCrCo},
  author={Niu, C and Zaddach, AJ and Oni, AA and Sang, X and Hurt, JW and LeBeau, JM and Koch, CC and Irving, DL},
  journal={Applied Physics Letters},
  volume={106},
  number={16},
  year={2015},
  publisher={AIP Publishing}
}

@article{wrobel2015phase,
  title={Phase stability of ternary fcc and bcc Fe-Cr-Ni alloys},
  author={Wr{\'o}bel, Jan S and Nguyen-Manh, Duc and Lavrentiev, Mikhail Yu and Muzyk, Marek and Dudarev, Sergei L},
  journal={Physical Review B},
  volume={91},
  number={2},
  pages={024108},
  year={2015},
  publisher={APS}
}

@article{cowley1965shortrange,
  author = {Cowley, J. M.},
  month = {05},
  pages = {A1384-A1389},
  title = {Short-Range Order and Long-Range Order Parameters},
  volume = {138},
  year = {1965},
  journal = {Physical Review}
}

@article{wolverton1994cluster,
  title={Cluster expansions of alloy energetics in ternary intermetallics},
  author={Wolverton, Chris and de Fontaine, Didier},
  journal={Physical Review B},
  volume={49},
  number={13},
  pages={8627},
  year={1994},
  publisher={APS}
}

@article{sofronis1995mechanics,
  title={Mechanics of the hydrogendashdislocationdashimpurity interactions—I. Increasing shear modulus},
  author={Sofronis, Petros and Birnbaum, Howard K},
  journal={Journal of the Mechanics and Physics of Solids},
  volume={43},
  number={1},
  pages={49--90},
  year={1995},
  publisher={Elsevier}
}

@article{sofronis2001hydrogen,
  title={Hydrogen induced shear localization of the plastic flow in metals and alloys},
  author={Sofronis, Petros and Liang, Yueming and Aravas, Nikolaos},
  journal={European Journal of Mechanics-A/Solids},
  volume={20},
  number={6},
  pages={857--872},
  year={2001},
  publisher={Elsevier}
}

@article{kim2019role,
  title={The role of hydrogen in hydrogen embrittlement of metals: the case of stainless steel},
  author={Kim, Young Suk and Kim, Sung Soo and Choe, Byung Hak},
  journal={Metals},
  volume={9},
  number={4},
  pages={406},
  year={2019},
  publisher={Multidisciplinary Digital Publishing Institute}
}

@article{flanagan1986hydrogen,
  title={Hydrogen induced disorder-order transition in Pd3Mn},
  author={Flanagan, Ted B and Craft, AP and Kuji, T and Baba, K and Sakamoto, Y},
  journal={Scripta metallurgica},
  volume={20},
  number={12},
  pages={1745--1750},
  year={1986},
  publisher={Elsevier}
}

@article{kim2015effect,
  title={Effect of short-range ordering on stress corrosion cracking susceptibility of Alloy 600 studied by electron and neutron diffraction},
  author={Kim, Young Suk and Maeng, Wan Young and Kim, Sung Soo},
  journal={Acta Materialia},
  volume={83},
  pages={507--515},
  year={2015},
  publisher={Elsevier}
}

@article{gerold1989origin,
  title={On the origin of planar slip in fcc alloys},
  author={Gerold, V and Karnthaler, HP},
  journal={Acta Metallurgica},
  volume={37},
  number={8},
  pages={2177--2183},
  year={1989},
  publisher={Elsevier}
}

@article{li2019strengthening,
  title={Strengthening in multi-principal element alloys with local-chemical-order roughened dislocation pathways},
  author={Li, Qing-Jie and Sheng, Howard and Ma, Evan},
  journal={Nature communications},
  volume={10},
  number={1},
  pages={1--11},
  year={2019},
  publisher={Nature Publishing Group}
}

@article{ding2018tunable,
  title={Tunable stacking fault energies by tailoring local chemical order in CrCoNi medium-entropy alloys},
  author={Ding, Jun and Yu, Qin and Asta, Mark and Ritchie, Robert O},
  journal={Proceedings of the National Academy of Sciences},
  volume={115},
  number={36},
  pages={8919--8924},
  year={2018},
  publisher={National Acad Sciences}
}

@article{zhang2020short,
  title={Short-range order and its impact on the CrCoNi medium-entropy alloy},
  author={Zhang, Ruopeng and Zhao, Shiteng and Ding, Jun and Chong, Yan and Jia, Tao and Ophus, Colin and Asta, Mark and Ritchie, Robert O and Minor, Andrew M},
  journal={Nature},
  volume={581},
  number={7808},
  pages={283--287},
  year={2020},
  publisher={Nature Publishing Group}
}

@article{san2008effects,
  title={Effects of alloy composition and strain hardening on tensile fracture of hydrogen-precharged type 316 stainless steels},
  author={San Marchi, C and Somerday, BP and Tang, X and Schiroky, GH},
  journal={International Journal of Hydrogen Energy},
  volume={33},
  number={2},
  pages={889--904},
  year={2008},
  publisher={Elsevier}
}

@article{klueh1988manganese,
  title={Manganese as an austenite stabilizer in Fe-Cr-Mn-C steels},
  author={Klueh, RL and Maziasz, PJ and Lee, EH},
  journal={Materials Science and Engineering: A},
  volume={102},
  number={1},
  pages={115--124},
  year={1988},
  publisher={Elsevier}
}

@article{vitos2006alloying,
  title={Alloying effects on the stacking fault energy in austenitic stainless steels from first-principles theory},
  author={Vitos, Levente and Nilsson, J-O and Johansson, B{\"o}rje},
  journal={Acta Materialia},
  volume={54},
  number={14},
  pages={3821--3826},
  year={2006},
  publisher={Elsevier}
}

@article{medvedeva2014first,
  title={First-principles study of Mn, Al and C distribution and their effect on stacking fault energies in fcc Fe},
  author={Medvedeva, NI and Park, MS and Van Aken, David C and Medvedeva, Julia E},
  journal={Journal of Alloys and Compounds},
  volume={582},
  pages={475--482},
  year={2014},
  publisher={Elsevier}
}

@article{jackson1984high,
  title={High temperature oxidation of iron-manganese-aluminum based alloys},
  author={Jackson, PRS and Wallwork, GR},
  journal={Oxidation of metals},
  volume={21},
  number={3},
  pages={135--170},
  year={1984},
  publisher={Springer}
}

@incollection{odegard1976effect,
  title={Effect of hydrogen on the mechanical behavior of nitrogen strengthened stainless steel},
  author={Odegard, BC and Brooks, JA and West, AJ},
  booktitle={Effect of hydrogen on behavior of materials},
  year={1976}
}

@article{zhang2008effect,
  title={Effect of nickel equivalent on hydrogen gas embrittlement of austenitic stainless steels based on type 316 at low temperatures},
  author={Zhang, Lin and Wen, Mao and Imade, Masaaki and Fukuyama, Seiji and Yokogawa, Kiyoshi},
  journal={Acta Materialia},
  volume={56},
  number={14},
  pages={3414--3421},
  year={2008},
  publisher={Elsevier}
}

@inproceedings{ferreira1996influence,
  title={Influence of hydrogen on the stacking-fault energy of an austenitic stainless steel},
  author={Ferreira, PJ and Robertson, Ian M and Birnbaum, HK},
  booktitle={Materials Science Forum},
  volume={207},
  year={1996}
}

@article{symons1997hydrogen,
  title={Hydrogen embrittlement of Ni-Cr-Fe alloys},
  author={Symons, DM},
  journal={Metallurgical and Materials transactions A},
  volume={28},
  number={3},
  pages={655--663},
  year={1997},
  publisher={Springer}
}

@article{Sanchez1984,
  title={Generalized cluster description of multicomponent systems},
  author={Sanchez, Juan M and Ducastelle, Francois and Gratias, Denis},
  journal={Physica A: Statistical Mechanics and its Applications},
  volume={128},
  number={1-2},
  pages={334--350},
  year={1984},
  publisher={Elsevier}
}

@article{birnbaum1994hydrogen,
  author = {Birnbaum, H.K. and Sofronis, P.},
  month = {03},
  pages = {191-202},
  title = {Hydrogen-enhanced localized plasticity—a mechanism for hydrogen-related fracture},
  volume = {176},
  year = {1994},
  journal = {Materials Science and Engineering: A}
}

\end{document}